# Quantum Mechanics of Insitu Synthesis of Inorganic Nanoparticles with in Anionic Microgels

Mirza Wasif Baig; Muhammad Siddiq

*Polymer Chemistry Laboratory 40;*

*Department of Chemistry, Quaid-i-Azam University, 45320, Islamabad, Pakistan.*

E-mail: mylifeischemistry.wasif@gmail.com

## Abstract:

In this work, we discuss the quantum mechanics of many-body systems i.e. hybrid microgel consisting of negatively charged anionic microgels possessing thick sheath of water molecules solvating its protruding anionic moieties and nanoparticle captivated with in it. Thermodynamic feasibility of synthesis of particular nanoparticle with in the microgel is dependent upon the magnitude of interaction between nanoparticle, water molecules and microgel relative to sum of magnitude of self-interaction between counterions and interaction between counterions and microgel. Nanoparticles synthesized with in the microgels have thick electronic cloud that oscillates under the influence of net interaction potential of charged anionic moieties and solvent water molecules. Hamiltonian describing energy of oscillating electronic cloud wrapped around nanoparticle is mathematically derived to be equal to product of integral of electron density and its position vector overall space multiplied with net electric force acting on the oscillating electronic cloud of nanoparticle i.e. $F \int \rho_n\{n\}\hat{n}dn$.

# Introduction:

Inorganic nanoparticles encompass vast portion of nanomaterials. Synthesis of nanoparticles constitutes a crucial aspect of nanochemistry. Several strategies have been employed for the synthesis of nanoparticles of different dimensionalities [1]. The controllable synthesis is attained in the presence of suitable surfactants, templates, capping agents such as polymers, ligands, and dendrimers [2]. Application of nanoparticles ranges from energy to medicine [3]. Optical properties of nanoparticles can be tuned by their size and shape which directly influence nanoscale excitons. Optical properties of NPS found their noticeable applications related to include light emitting devices, lasers, photovoltaics, detectors, and biolabels [4]. The application of nanoparticles in medicine has given birth to nanobiotechnology. Nanoparticles have ability to penetrate the cell wall and deliver drugs or biomolecules into living systems for a therapeutic purpose [5]. For last decade synthesis of nanoparticles in side polymer microgels have gained much attention. Template based synthesis of nanoparticles in the interior of the microspheres is an alternative and effective approach for the synthesis of semiconductor, metal and magnetic particles [6]. M. Karg and T. Hellweg have written a self-explanatory review about use of poly (N-isopropyl-acrylamide) as nanoreactor for synthesis of inorganic nanoparticles [7]. Use of anionic microgels as microreactor has been proved to be one of the most successful chemical methods for size and shape controlled synthesis of inorganic nanoparticles [8]. Microgels are synthesized by free radical emulsion polymerization of NIPAM with one or two appropriate monomers with anionic groups (normally carboxylic acid and sulphonic acid moieties), methylene bisacrylamide as cross linking agent. Microgel is three dimensional networks with meshes in it possessing protruding anionic moieties with in it that hold up water molecules and responsible for the swelling and deswelling of behavior of microgels [9]. Mesh present in typical microgel molecule can be shown in following figure,

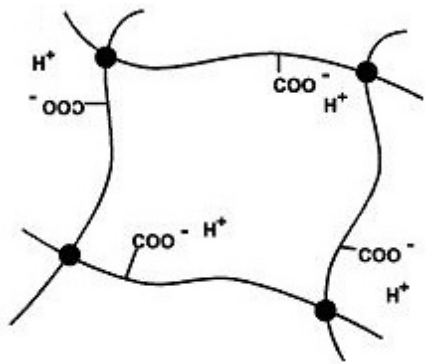

**Figure 1.** Structure of a mesh in an anionic microgel.

Microgels provide a very handy method for synthesis of nanoparticles at room temperature. Mesh present in microgels acts as microreactor for synthesis of nanoparticles. Size of mesh and number of charged anionic groups protruding in it together dictates the size and morphology of nanoparticles. pH of external media controls the ionization of anionic groups and also plays key role in dictating the size and shape of nanoparticles. In aqueous solution of microgels solution of metal salt of known concentration is added and stirred for a reasonable time to achieve complete homogeneity. Suitable reducing agent normally $NaBH_4$ is added that reduces the metal cations having positive reduction potential or having reduction potential less negative than reducing agent to neutral atoms. In other words thermodynamically feasible reduction can be carried out with in microgel. Atoms reduced coagulate to form nanoparticles whose size is dictated by above mentioned factors. Nanoparticles are of size exceeding then the size of mesh in which they are produced this result in their long lasting captivity with in the respective mesh of the microgel. Protruding anionic groups which initially attracted metal cations with in the mesh are now in position to repel the oscillating electronic cloud present at the surface of nanoparticle which is responsible for its optical and catalytic properties. This repulsion is overcome by development of thick sheath of water molecules that solvates the captivated nanoparticle. Solvation is in such a way that bipolar solvent molecules orient one of their partial positively charge end towards the electronic sheath of nanoparticle and other end towards the negatively charged end of anionic moieties of microgels thus stabilizing the charge repulsion and thus stabilizing the system. Microgels permanently captivate nanoparticles with in it and thick sheath of water molecules solvates them. Nanoparticle captivated with in the mesh of microgel can be shown in following figure,

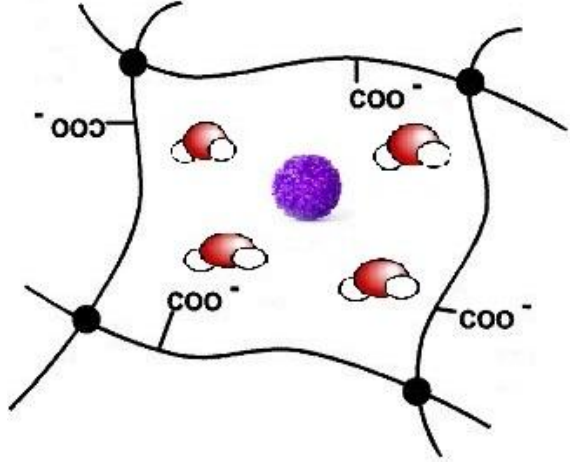

**Figure 2.** Structure of a mesh of hybrid anionic microgel captivating a nanoparticle.

## Theoretical Model:

According to best of our knowledge physics behind the insitu synthesis of nanoparticles with in the microgel has not been reported yet. Here we formulate quantum mechanics behind this insitu synthesis of nanoparticles with in the microgel. Let consider initially microgels to be dispersed in aqueous phase which is devoid of any other counter ions just possessing hydrogen ions produced from the ionization of anionic groups (normally carboxylic acid and sulphonic acid moieties) present in the microgels. Let assume microgels possessing $N_m^-$ anionic moieties giving up $N_{H^+}^+$ hydrogen ions. On adding solution of metal salt MX in microgel solution produces $N_M^+$ cations and anions $N_X^-$. Electrical neutrality is achieved with in microgel by $N_m^- = N_M^+$ and outside microgel by $N_{H^+}^+ = N_X^-$. Interaction of microgels and counter metal cations can be mathematically modeled by following total Hamiltonian [10],

$$H = H_m(\{x\}) + H_c(\{y\}) + H_{mc}(\{x\},\{y\}) \qquad (1)$$

Hamiltonian $H_m$ for microgel is described in terms of kinetic energy and potential energy terms. First term is kinetic energy term and second term is potential energy term which explains potential of interaction among different microgel negatively charged moieties described by coordinates $x$. Hamiltonian for microgel is written as [11-12],

$$H_m = K_m + \frac{1}{2}\sum_{i \neq j=1}^{N_m} v(|x_i - x_j|) \qquad (2)$$

Hamiltonian $H_c$ for counter ions is described in terms of kinetic energy and potential energy terms. First term is kinetic energy term and second term is potential energy term which explains potential of interaction among different counterions described by coordinates $y$. Hamiltonian for counter ion is written as [10],

$$H_c = K_c + \frac{1}{2}\sum_{i \neq j=1}^{N_c} v(|y_i - y_j|) \qquad (3)$$

Hamiltonian $H_{mc}$ is written in terms of potential energy term and describes potential produced due to attractive interaction between counter ions and negative moieties of microgels and is mathematically expressed as,

$$H_{mc} = \sum_{i \neq j=1}^{N_m} \sum_{i \neq j=1}^{N_c} v(|x_i - y_j|) \qquad (4)$$

Hamiltonian $H_{mc}$ in terms of density operators of microgel and density operators of counter ion can be written in following integral form;

$$H_{mc} = \int \rho_m\{x\}dx \int \rho_c\{y\}dy \qquad (5)$$

Where density operators for microgel and counter ion can be defined as [12];

$$\rho_m\{x\} = \sum_{j=1}^{N_m} \delta(x - x_j) \qquad (6)$$

$$\rho_c\{x\} = \sum_{j=1}^{N_c} \delta(y - y_j) \qquad (7)$$

The mixture of microgels and counterions can be expressed by one component Hamiltonian by tracing coordinates of counter ion and partition function for counterions can be expressed as;

$$Z_c = exp[-H_c + H_{mc}/k_B T] \qquad (8)$$

Free energy associated with counterions can be expressed in terms of partition function as;

$$G_c = -k_B T In(exp[-H_c + H_{mc}/k_B T]) \qquad (9)$$

When suitable reducing agent is added in solution of microgels and counterions; they get reduced and neutral atoms start agglomerating to form nanoparticles whose size is greater than the size of mesh and thus they are captivated with in the respective mesh where reduction is carried out. Mostly single nanoparticle is formed with in one mesh of microgel. Nanoparticle formed is aggregation of hundreds to thousands of atoms wrapped with electronic sheath. Let on reduction of metal cations captivated with in the meshes of microgels and produce metal nanoparticle $M_P$ with in it. Now negatively charged groups normally carboxylic acid groups or sulphonic acid groups will not own this neutral nanoparticle wrapped with negatively charged electronic sheath. Thus thick sheath of water molecules is sandwiched between electronic sheath of nanoparticle and protruding negative charge moieties with in the mesh captivating the nanoparticle. Interaction between microgel negatively charged groups, strictly hold water sheath and nanoparticles can be mathematically modeled by following total Hamiltonian,

$$H = H_n(\{n\}) + H_m(\{x\}) + H_w(\{z\}) + H_{wn}(\{z\},\{n\}) + H_{wm}(\{z\},\{x\}) + H_{mn}(\{x\},\{n\}) \tag{10}$$

Hamiltonian of nanoparticle described by coordinate $n$ is just composed of kinetic energy term as it can be regarded as a single particle captivated with in a particular mesh of microgel and thus it can be mathematically expressed as;

$$H_n(\{n\}) = K_n \tag{11}$$

Above Hamiltonian describing kinetic energy of nanoparticle can be approximated as zero because nanoparticle captivated in a mesh is composed of hundreds of atoms that hardly execute any translational motion but simply rotational motion which is too slow and ignorable and this term can be discarded i.e. $H_n(\{n\}) \cong 0$.

Hamiltonian $H_w$ for thick sheath of water molecules sandwiched between the negatively charged moieties of microgel and nanoparticle is described in terms of kinetic energy and potential energy terms. First term is kinetic energy term and second term is potential energy term which explains potential of interaction among different water molecules of thick sheath embedded between neutral nanoparticles and negatively charged moieties of the mesh described by coordinates $z$. Hamiltonian for microgel is written as,

$$H_w = K_w + \frac{1}{2}\sum_{i \neq j=1}^{N_w} v(|z_i - z_j|) \tag{12}$$

Hamiltonian $H_{wm}$ is written in terms of potential energy term and describes potential produced due to attractive interaction between water molecules and negatively charged moieties of microgel and is mathematically expressed as,

$$H_{wm} = \sum_{i \neq j=1}^{N_w} \sum_{i \neq j=1}^{N_m} v(|z_i - x_j|) \tag{13}$$

Hamiltonian $H_{wm}$ in terms of density operators of water and density operators of microgel can be written in following integral form;

$$H_{wm} = \int \rho_w\{z\}dz \int \rho_m\{y\}dy \tag{14}$$

Where density operators for water and nanoparticle can be defined as;

$$\rho_w\{z\} = \sum_{j=1}^{N_w} \delta(z - z_j) \tag{15}$$

$$\rho_m\{x\} = \sum_{j=1}^{N_m} \delta(x - x_j) \tag{16}$$

Hamiltonian $H_{wn}$ can be expressed only in terms of potential energy term produced due to attractive interaction between water molecules and neutral nanoparticle present with in the microgel.

$$H_{wn} = \int \rho_n\{n\}v(n)dn \tag{17}$$

Potential energy originates due to attractive interaction between water molecules and neutral nanoparticle is between positively charged ends of water molecules with the electronic sheath wrapped around the nanoparticle. Oscillating electronic sheath wrapped around several bounded metallic nuclei that comprises the nanoparticle can be treated as a vibrating body. Electronic sheath oscillates under the influence of electric field of these several bounded nuclei with a measurable force constant $k$. Nuclei comprising nanoparticles can be approximated stationary with respect to oscillating electronic cloud this approximation is on the behalf of difference of mass of electronic cloud and nuclei. When thick sheath of water molecules surrounds the captivated nanoparticle then electronic sheath comes under the influence of electric field generated by oriented positively charged poles of thick sheath of water molecules. Net electric field $\xi_w$ generated by oriented positively charged poles of water molecules polarize the electronic cloud of nanoparticle and induces the dipole moment in it. Thus Hamiltonian $H_{wc}$ in terms of dipole moment induced by electric field of water molecules can be mathematically expressed as,

$$H_{wn} = -\xi_w \int \rho_n\{n\}\hat{\mu}(n)dn \quad (18)$$

Dipole moment operator is product of position vector of electronic sheath and total charge it possess and thus Hamiltonian $H_{wc}$ becomes,

$$H_{wn} = -\xi_w Pe \int \rho_n\{n\}\hat{n}dn \quad (19)$$

Hamiltonian $H_{mn}$ is expressed in terms of potential energy term and describes potential produced due to interaction between negatively charged moieties of microgels and neutral nanoparticle.

$$H_{mn} = \int \rho_n\{n\}v(n)dn \quad (20)$$

Interaction between nanoparticles and microgel can be totally explained by interaction between protruding negatively charged anionic moieties (carboxylic acid groups and sulphonic acid groups) and wrapped electronic sheath along the periphery of nanoparticle. Electric field of anionic moieties $\xi_a$ also polarizes the electronic sheath by repulsive interaction inducing and inducing an electric field opposite to that induced by electric field of sheath of solvating water molecules. Thus $H_{mn}$ is mathematically expressed as,

$$H_{mn} = \xi_a \int \rho_n\{n\}\hat{\mu}(n)dn \quad (21)$$

Dipole moment operator is product of position vector of electronic sheath and total charge it possess and thus Hamiltonian $H_{wc}$ becomes,

$$H_{mn} = \xi_a Pe \int \rho_n\{n\}ndn \quad (22)$$

The mixture of microgels, water and nanoparticle can be expressed by one component Hamiltonian by tracing coordinate of nanoparticle and then partition function for nanoparticle can be expressed as;

$$Z_n = exp[-H_{wn} + H_{mn}/k_BT] \quad (23)$$

Free energy associated with nanoparticle can be expressed in terms of partition function as;

$$G_n = -k_B T ln(exp[-H_{wn} + H_{mn}/k_BT]) \quad (24)$$

When reduction is carried out metal counterions present with in the meshes of microgels get reduced to nanoparticle insitu and free energy change accompanied with it can be expressed as;

$$\Delta G_{insitured} = G_n - G_c \qquad (25)$$

Substituting free energies of counter ions and nanoparticles gives following relation;

$$\Delta G_{insitured} = -k_B T \ln(\exp[-H_{wn} + H_{mn}/k_B T]) + k_B T \ln(\exp[-H_c + H_{mc}/k_B T]) \qquad (26)$$

Then insitu reduction will be spontaneous and exergonic that is;

$$\Delta G_{insitured} < 0 \qquad (27)$$

From Eq. (26) it can be mathematically interpreted that $\Delta G_{insitured} < 0$ if;

$$H_{wn} + H_{mn} > H_c + H_{mc} \qquad (28)$$

Net Hamiltonian can also be written for describing interaction between anionic moieties of microgels and thick sheath of solvated water molecules and can be mathematically expressed as;

$$H_{net} = H_{wn} + H_{mn} = (\xi_a - \xi_w) Pe \int \rho_n\{n\} \hat{n} dn \qquad (29)$$

From Eq. (29) it is evident that if electric field of water molecules $\xi_w$ is greater in magnitude than that of electric field generated by anionic moieties $\xi_a$ then value of net Hamiltonian will be attractive in nature and captivity of nanoparticle with in a mesh will be energetically feasible. Electric force acting on the electronic sheath of nanoparticle $\mathcal{F}$ can be set equal to product of net electric field with in the mesh and charge possess by electronic sheath of nano particle i.e. $(\xi_a - \xi_w) Pe$ and thus $H_{net}$ transform as;

$$H_{net} = H_{wn} + H_{mn} = \mathcal{F} \int \rho_n\{n\} \hat{n} dn \qquad (30)$$

## Discussion:

Initially water molecules are being attracted with in a mesh by force of attraction with anionic moieties and counter ions with in the mesh after carrying out reduction with in mesh of microgel they also get engaged in developing interaction with electronic cloud of the nanoparticle and overcomes repulsive interaction between anionic moieties of gel and oscillating electric cloud of nanoparticle. Interaction between solvent water molecules and charge cloud of nanoparticle results in entrance of more water molecules inside the mesh followed by development of thick sheath of solvated water molecules with in the mesh around the captivated nanoparticle. Electric fields are set up across the electronic cloud of nanoparticle by protruding anionic moieties of microgel and solvent water molecules with in the microgel that governs its swelling and deswelling behavior. Now electronic sheath wrapping up the nanoparticle oscillates in presence of an additional electric field set up by anionic moieties of microgel and solvent water molecules. Net electric field appreciably polarizes the electronic cloud of nanoparticles inducing a dipole moment in it. Hamiltonian describing interaction between nanoparticle, microgel and solvent water molecules is defined in terms of this induced dipole moment set up in the presence of net electric field. Magnitude of this interaction is directly related to the polarization behavior of nature of element under study. So soft metals that provide enough flexibility to their electronic cloud to get polarized under the influence of applied electric field and could hold the induce dipole for a pretty measurable time. Thus metal atoms with low ionization potential and high polarizability that can show affinity for sheath of water molecules solvating them leads to thermodynamically feasible captivity of nanoparticles. Thermodynamics of insitu reduction is totally dictated by this interaction. If eigen value of Hamiltonian describing interaction between induced dipole moment in electronic cloud of nanoparticle is greater than eigen value of Hamiltonians defining interaction between counter ions, water molecules and microgels then reduction becomes thermodynamically feasible. Magnitude of free energy is evaluated by difference of respective partition function. If electric field of solvated water particles is greater than electric field of anionic groups of microgels then electronic cloud develop attractive interaction and dipole induced governs it various behavior including red shift in its surface plasmon band. This interaction between induced dipole of electric cloud and net electric field also contributes majorly towards encapsulation of nanoparticles with in the microgel. Encapsulation of nanoparticles with in a microgel is so strong that it strongly tunes its catalytic properties, as it hampers the upcoming substrate molecules toward nanoparticle and thus also reducing its catalytic activity [13].

**Conclusion:**

Exergonicity of the insitu reduction of metal cations with in anionic microgel is dictated by the strength of interaction of electronic cloud of nanoparticle with net electric field generated by protruding anionic moieties, water molecules present in the mesh of microgel. Extent of polarization of electronic cloud is very crucial for the thermodynamics feasibility of synthesis of nanoparticle with in a microgel. Energy of electronic cloud of nanoparticle captivated with in a microgel can be obtained by product of integral of electron density and its position vector overall space multiplied with net electric force acting on it i.e. $F \int \rho_n\{n\}\hat{n}dn$. Magnitude of this integral defines the fate of synthesis of nanoparticle if its magnitude is greater than the magnitude of Hamiltonians describing self-interaction between counter ions and interaction between counter ions and microgel then reduction exergonic.